\documentclass[review]{elsarticle}
\UseRawInputEncoding  
\usepackage{amsmath}
\usepackage{amssymb}
\usepackage{amsthm}
\usepackage{psfrag}  
\usepackage{amsfonts,amsmath,enumerate,graphicx,enumerate}
\usepackage{lineno}
\modulolinenumbers[5]

\journal{Journal of \LaTeX\ Templates}

\allowdisplaybreaks

\newcommand{\Lower}[1]{\smash{\lower 1.3ex  \hbox{#1}}} 

\def\globalmin{\mathop{\rm global \; min}}

\begin{document}

\begin{frontmatter}

\title{Mathematical modelling of zika virus in Brazil}



\author[mymainaddress1]{Sa\'ul E. Buitrago Boret\corref{mycorrespondingauthor}}
\cortext[mycorrespondingauthor]{Corresponding author}
\ead{sbutrago@usb.ve}

\author[mymainaddress2]{Ren\'e Escalante}
\ead{rene.escalante@uah.es}

\author[mymainaddress1]{Minaya Villasana}
\ead{mvillasa@usb.ve}

\address[mymainaddress1]{Dpto. C\'omputo Cient\'{\i}fico y Estad\'{\i}stica, Universidad Sim\'on
Bol\'{\i}var, Caracas, Venezuela}
\address[mymainaddress2]{Dpto. de F\'{\i}sica y Matem\'aticas, Universidad de Alcal\'a, Madrid, Spain}

\begin{abstract}
In this paper we study some deterministic mathematical models that seek
to explain the expansion of zika virus, as a viral epidemic, using
published data for Brazil.
SIR type models are proposed and validated using the epidemic data found,
considering several aspects in the spread of the disease.
Finally, we confirmed that the crucial epidemic parameter such as $R_0$
is consistent with those previously reported in the literature for other areas.
We also explored variations of the parameters within Brazil for different federal
entities.
We concluded that a parsimonious model that includes both human and vector
populations best describe the epidemic parameters.
\end{abstract}


\end{frontmatter}

\nolinenumbers

\section{Introduction} \label{Sec_Introduction}

The zika virus (ZIKV) was first isolated in
1947 from a sentinel rhesus monkey in the Zika forest in Uganda
(see~\cite{Dicketal1952}) and was
classified by sequence analysis into two genotypes, African and Asian
(see~\cite{Fayeal2013}).
In April 2007, a large epidemic of Asian genotype ZIKV was reported in Yap
Island and Guam, Micronesia.
Between 2013-2014 the Asian genotype caused epidemics reported in several
Pacific Islands, including French Polynesia, New Caledonia,
Cook Islands, Tahiti, and Easter Island.
(see~\cite{FariaEtAl2016} and \cite{BerkowitzEtAl2016}).

In a general review published in 2014 by Ioos et al.~\cite{Ioosal2014}
it was reported that the ZIKV infection caused two major epidemics
in Pacific previously naive territories, in less than a decade.
This emergent arbovirosis transmitted by mosquitoes of the Aedesgenus has
a high potential for spreading in countries where the vector is present.
In March 2015, the first endogenous transmission of Zika virus in
Brazil was reported. Subsequent studies in mice suggest that the virus could attack the adult brain
as well (see \cite{Bernardo2016} and \cite{LiHetal2016}).

\begin{figure}[htp!]
   \begin{center}
   \includegraphics[width=11.0cm]{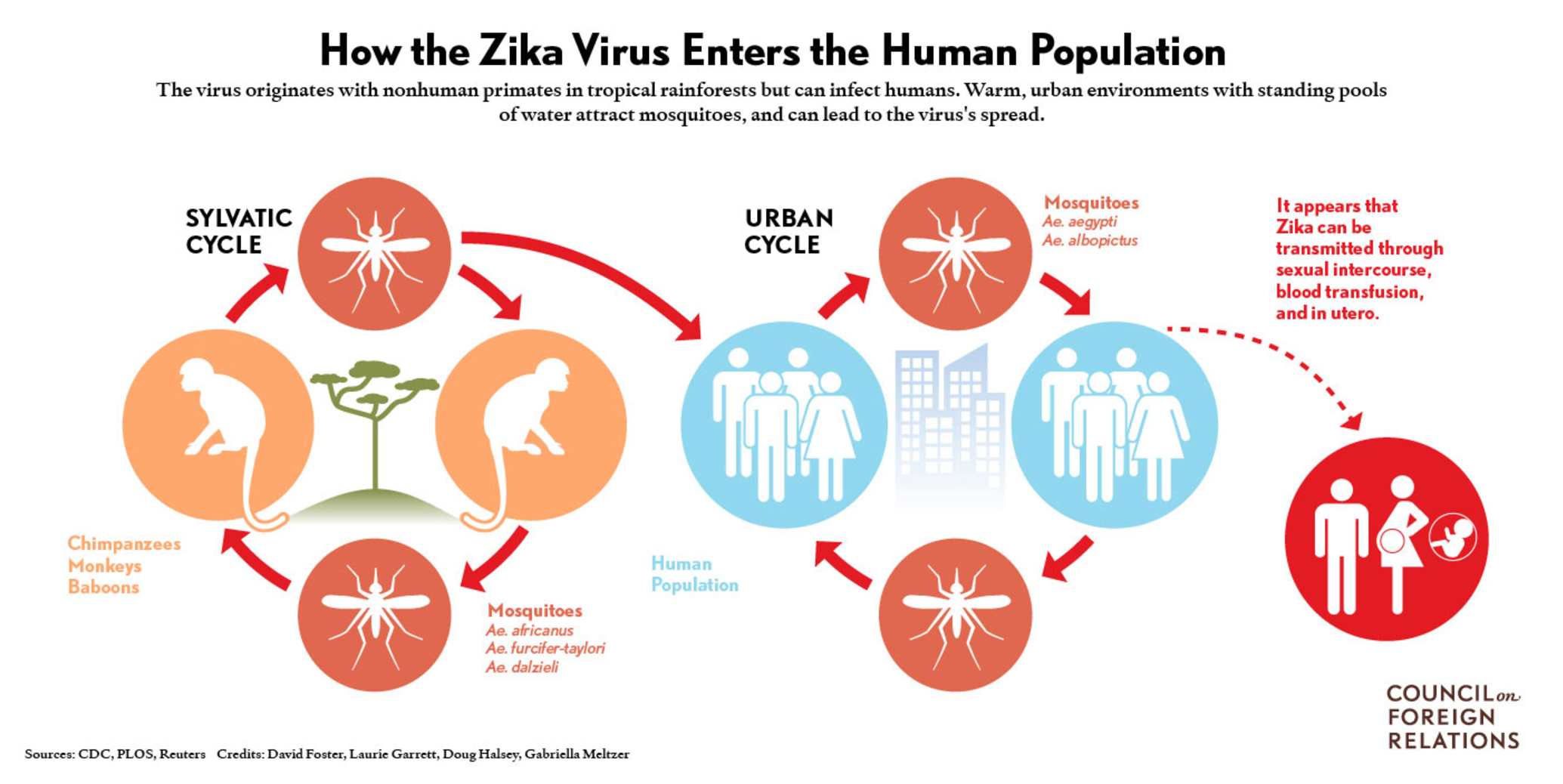}
   \vspace*{-2mm}
   \caption {How the zika virus enters the human population.
     Source: CDC, PLOS, Reuters; Credits: David Foster, Laurie Garrett, Doug Halsey
       and Gabriela Meltzer.}
   \label{zika-enters-humans}
   \end{center}
\end{figure}

ZIKV infection, together with denge and chikungunya, are one of the leading causes
of illness in the tropics and subtropics, where it inflicts substantial health,
economic and social burdens.
Humans are infected with zika virus by the bite of an infective female mosquito
Aedes aegypti, the principal vector of zika. Once a person gets bitten by an infective
mosquito, the virus undergoes an incubation period of about 3 to 12 days, after which
the person enters the acute phase of infection. The acute phase can be as short as 2
days and as long as 7 days. If other female Aedes aegypti mosquitoes bites the ill
person during this acute phase, those mosquitoes may become infected and subsequently
begin the transmission cycle anew.
Fig.~\ref{zika-enters-humans} (Source: CDC, PLOS, Reuters; Credits: David Foster,
Laurie Garrett, Doug Halsey and Gabriela Meltzer) shows graphically how zika
virus enters the human population. 

Many people infected with zika virus will not have symptoms or will only have
mild symptoms. The most common symptoms of zika are fever, rash, joint pain,
conjuntivitis (red eyes), muscle pain and headache.
Zika is usually mild with symptoms lasting for several days to a week.
People usually do not get sick enough to go to the hospital, and they very rarely
die of zika. For this reason, many people might not realize they have been infected.
Symptoms of zika are similar to other viruses spread through mosquito bites,
like dengue and chikungunya.

There is scientific consensus that zika virus is a cause of microcephaly and
Guillain-Barr\'e syndrome (see~\cite{WHOplan2016}). 
Links to other neurological complications are also being investigated.
Sexual transmission of zika virus is also possible. Other modes of transmission
such as blood transfusion are being investigated.

The foundations of the entire approach to epidemiology,
based on compartmental models, were laid by public health physicians such as
Sir R.A. Ross \cite{Ross1911}, W.H. Hamer, A.G. McKendrick, and W.O. Kermack
\cite{KermackEtAl1927,KermackEtAl1932,KermackEtAl1933} between 1900 and
1935, along with important contributions from a statistical perspective by
J. Brownlee \cite{Brownlee1916,FarewellEtAl2013}.

Mathematical models have been extensively used to study the dynamics of infectious
diseases at population level.
Most continuous time models are in the form of ordinary differential equations
(ODEs). Such ODE models assume that the population is well mixed, and
the transmission is instantaneous (see~\cite{BrauerEtAl2012, BrauerEtAl2015}).

Mathematical modelling is typically the only way to
examine the possible impact of different release and control scenarios.
Questions that can be addressed are, for instance,
what fraction of the population should be quarantined and/or vaccinated?
How fast can control measures to be implemented?, etc.

The basic reproductive ratio (see~\cite{Diekmannetal1990, Diekmannetal2010} and
\cite{Heffernanetal2005}), $R_0$, is defined as the expected number of
secondary infections arising from a single individual during his or her
entire infectious period, in a population of susceptibles.
$R_0$ often serves as a threshold parameter that predicts whether an infection
will spread.
Determinig $R_0$ is vital to understand and characterize the dynamics
of the disease. However this crucial parameter is model dependent.
We study and calculate $R_0$ using different modelling perspectives that can
allow us to draw better conclusions on its validity and range.
Also, this parameter is investigated at different granularity levels:
country wide and state wide.

This paper, based on the preliminary work by Buitrago {\it et~al.}
\cite{BuiEscVil2016}, is organized as follows:
section~\ref{Sec_Math_Model} outlines the mathematical models used,
section~\ref{Sec_R0} briefly describe details around the basic reproductive
ratio $R_0$ for the models developed in section~\ref{Sec_Math_Model},
section~\ref{Sec_Results} provides information about the data used and
in the last section, section~\ref{Sec_Discussion},
a discussion of the methodologies and their application to the data
sets are summerized.

\section{Mathematical Modelling} \label{Sec_Math_Model}

We formulate our descriptions as compartmental models, with the population under
study being divided into compartments and with assumptions about the nature
and time rate of transfer from one compartment to another.


In formulating models in terms of the derivatives of the sizes of
each compartment we are also assuming that the number of members in a
compartment is a differentiable function of time. This assumption is
plausible once a disease outbreak has become established but is not valid
at the beginning of a disease outbreak when there are only a few infectives.

In this work we describe models for epidemics, acting on a sufficiently rapid
time scale that demographic effects, such as births, natural deaths,
maintain a constant level of the overoll population, and migration may be ignored.

All the models considered in this work satisfy the following assumptions:
\begin{itemize}
\item There is homogeneous mixing, which means that individuals of the
population make contact at random and do not mix mostly in a smaller subgroup.
\item The disease is novel, so no vaccination is available and or applied.
\item Any recovered person has permanent immunity or least considered
as such within the time-frame of the disease.
\item The population size is constant for the models.
\end{itemize}

\subsection{SIR model} \label{SSec_SIR_model}
Consider a population in which a small number of its members suffer from an
infectious disease that can be transmitted to other members of the same population.
The objective we are pursuing now is to determine what proportion of the total
population will be infected and for how long, using a mathematical model that
incorporates into their structure the transmission mechanisms that we consider
important.

In order to model such an epidemic we divide the population being studied into
three classes labeled $S$, $I$, and $R$.

\begin{figure}[htp!]
   \begin{center}
   \includegraphics[width=9.0cm]{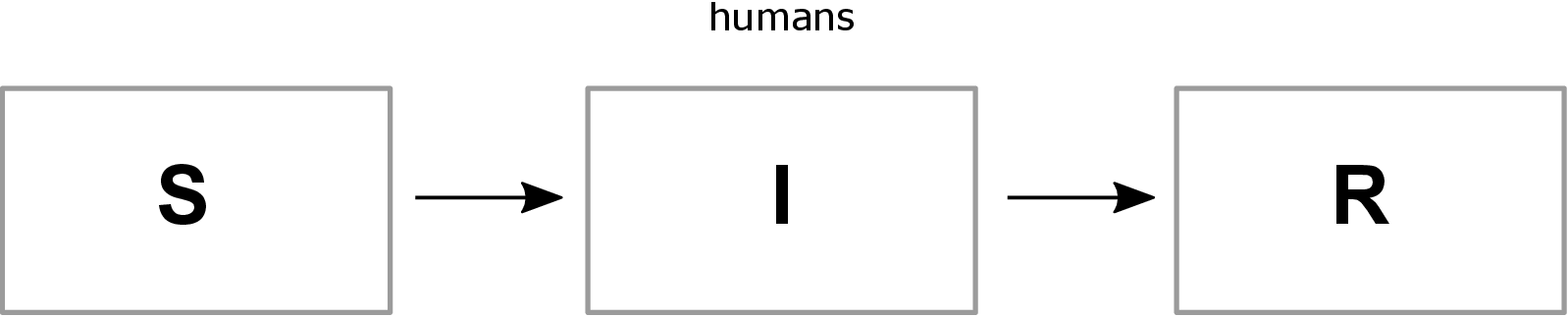}
   \caption {Structure of the SIR model.}
   \label{SIR_model}
   \end{center}
\end{figure}

Let $S(t)$ denote the number of individuals who
are susceptible to the disease, which can acquire the infection through contacts
with infectious, that is, who are not (yet) infected at time $t$.
$I(t)$ denotes the number of infected individuals, assumed infectious and able to
spread the disease by contact with susceptibles.
$R(t)$ denotes the number of individuals who
have been infected and then removed from the possibility of being infected again or
of spreading infection (see Fig.~\ref{SIR_model}).

Removal is carried out through isolation from the rest of the
population, through immunization against infection, through recovery from the disease
with full immunity against reinfection, or through death caused by the disease.
These characterizations of removed members are different from an epidemiological
perspective but are often equivalent from a modelling point of view that takes into
account only the state of an individual with respect to the disease.

We will use the terminology SIR to describe a disease that confers immunity
against reinfection, to indicate that the passage of individuals is from the susceptible
class $S$ to the infective class $I$ to the removed class $R$.
The mathemaical model is:

\begin{eqnarray}
\frac{dS}{dt}(t) & = & \mu N -\beta S(t) I(t) /N - \mu S(t), \label{dsdt} \\
\frac{dI}{dt}(t)  & = & \beta S(t) I(t) /N - \gamma I(t) - \mu I(t), \label{didt} \\
\frac{dR}{dt}(t) & = & \gamma I(t) - \mu R(t), \label{drdt} \\
\frac{dC_i}{dt} & = & p \beta S(t) I(t), \label{dcidt}
\end{eqnarray}
with the initial conditions in $t = t_0$
\begin{eqnarray}
S(t_0) & = & S_0 > 0, \label{inicialS} \nonumber \\
I(t_0) & = & I_0 > 0, \label{inicialI} \nonumber \\
R(t_0) & = & R_0 > 0, \label{inicialR} \nonumber
\end{eqnarray}
where $S_0$, $I_0$ and $R_0$ are, respectively, the initial number of suceptible,
infected and recovered people with $\beta$, $\gamma$ and $\mu$ positive constants.
As is often the case, not all infectives are symptomatic, specially in the Zika
virus, and thus not all cases are reported as such making the determination
of the real number of infectives a difficult task.
In equation~\ref{dcidt}, $C_i$ accounts for the cumulative infectives and is
a smooth monotone function that is used for identification purposes.
The parameter $p$ is a proportion of the infectives that are reported.
$\beta$ is the transmission rate from mosquitoes to humans,
$\gamma$ is the per capita rate of recovery in humans such that
$1/\gamma$ is the mean infectious period for humans,
$\mu$ is the per capita rate of mortality in humans such that $1/\mu$
is the life expectancy of humans, and $N$ is the human population size.

\subsection{SIR/SI model} \label{SSec_SIRSI_model}
This model is an extension of the SIR model,
and has been used before in the study of the dynamics of dengue in
Thailand by Pandey {\it et~al.} ~\cite{Pandey2013}.

\begin{figure}[htp!]
   \begin{center}
   \includegraphics[width=9.0cm]{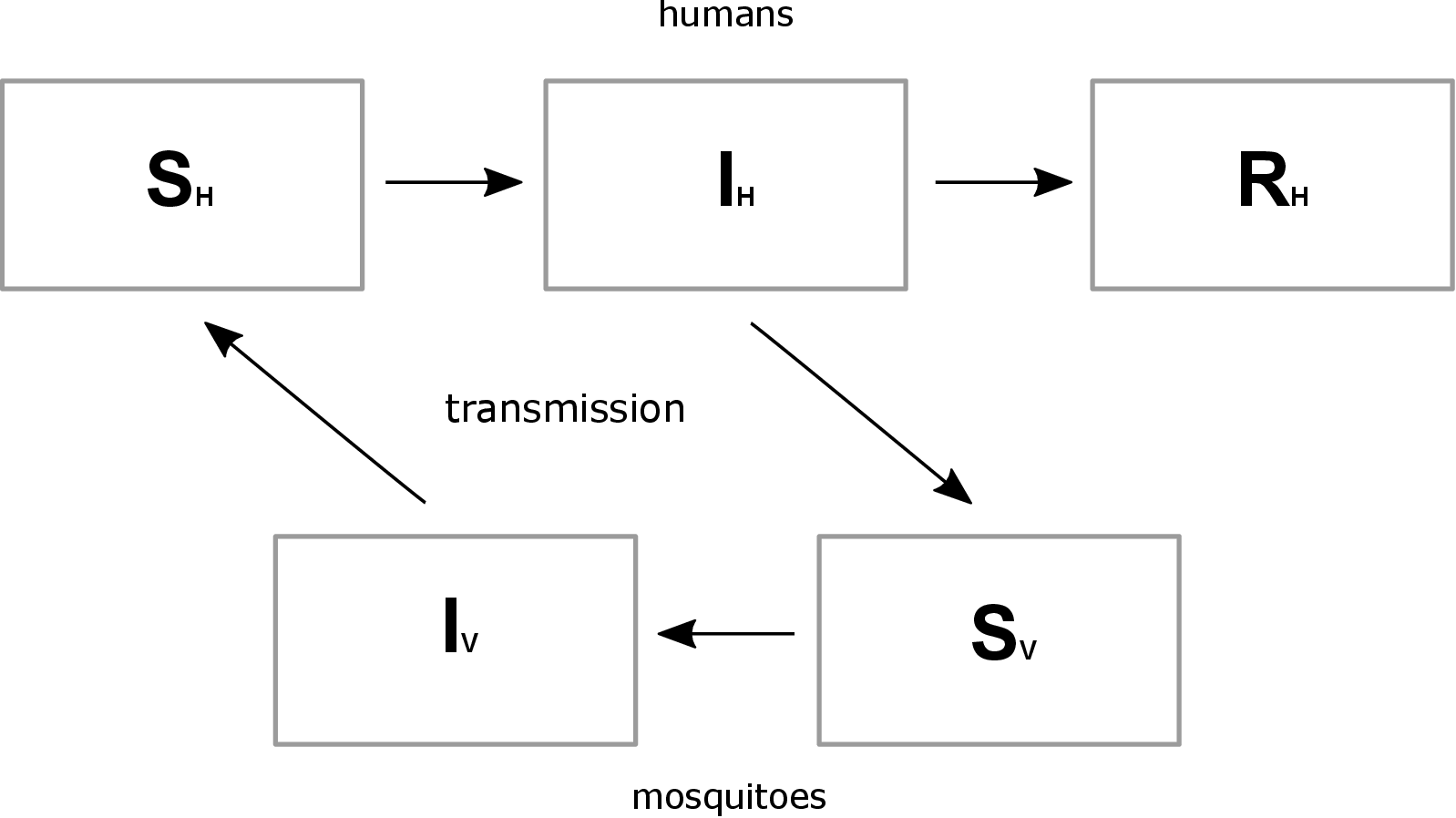}
   \caption {Structure of the SIR/SI model.}
   \label{SIRSI_model}
   \end{center}
\end{figure}

The population is divided into three classes for humans and two clases
for mosquitoes or vectors that transmit the disease.
$S_H$ represents the number of susceptible, $I_H$ the number of infectious, and
$R_H$ the number recovered individuals in the human sub-population.
Similarly, $S_v$ represents the proportion of mosquitoes currently susceptible,
and $I_v$ the proportion of infectious mosquitoes infectious (see Fig.~\ref{SIRSI_model}).

Mosquitoes are assumed to remain infectious for life.
$\beta_v$ is the transmission rate from humans to mosquitoes,
$\beta_H$ is the transmission rate from mosquitoes to humans,
while $1/\gamma_H$ and
$1/\mu_H$ are the mean infectious period and the mean lifespan of humans,
$1/\mu_v$ is the mean lifespan of mosquitoes.
$N_H$ stands for the human population size and $N_v$ is the mosquito
population size.
The mathematical model is written as follows:

\begin{eqnarray}
\frac{dS_H}{dt}(t) & = & \mu_H N_H -\beta_H S_H(t) I_v(t) /N_v - \mu_H S_H(t), \label{dsHdt2} \\
\frac{dI_H}{dt}(t)  & = & \beta_H S_H(t) I_v(t) /N_v - \gamma_H I_H(t) - \mu_H I_H(t), \label{diHdt2} \\
\frac{dR_H}{dt}(t)  & = &  \gamma_H I_H(t) - \mu_H R_H(t) \label{RH2} \\
\frac{dI_v}{dt}(t)  & = & \beta_v S_v(t) I_H(t) /N_H - \mu_v I_v(t), \label{divdt2} \\
\frac{dS_v}{dt}(t) & = & -\beta_v S_v(t) I_H(t) /N_H + \mu_v I_v(t), \label{Sv2} \\
\frac{dC_i}{dt} & = & p \beta_H S_H(t) I_v(t) / N_v, \label{dciHdt2}
\end{eqnarray}
with the initial conditions in $t = t_0$
\begin{eqnarray}
I_H(t_0) & = & I_{H0} > 0, \label{inicialIH2} \nonumber \\
R_H(t_0) & = & R_{H0} > 0, \label{inicialRH2} \nonumber \\
I_v(t_0) & = & I_{v0} > 0, \label{initialIv2} \nonumber
\end{eqnarray}
where $I_{H0}$, $R_{H0}$ and $I_{v0}$ are respectively, the initial number of
infected people, the initial number of recovered people, and the initial number
of infectious mosquitoes, respectively.
As in the previous model, we use the cummulative number of infectives for
identification purposes (equation~\ref{dciHdt2}).

Given the fact that the human population remains constant, one can express
$R_H$ in terms of the variables $S_H$ and $I_H$, i.e. $R_H = N_H - S_H -I_H$,
therefore equation \ref{RH2} can be discarded and we can reduce the
dimensionality of the system.
A similar argument is true for the case of $S_v$ in equation~\ref{Sv2},
i.e. $S_v = N_v - I_v$.

\subsection{SEIR/SEI model} \label{SSec_SEIRSEI_model}
\begin{figure}[htp!]
   \begin{center}
   \includegraphics[width=13.0cm]{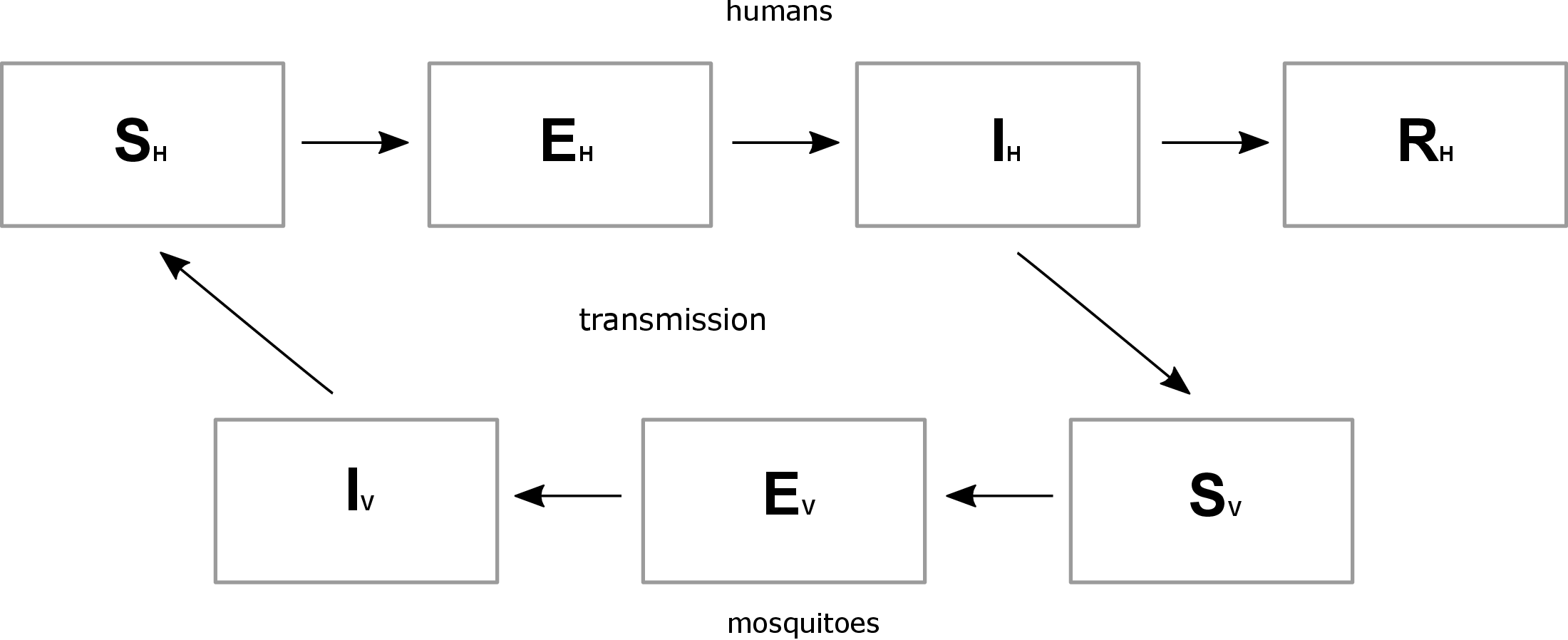}
   \caption {Structure of the SEIR/SEI model.}
   \label{SEIRSEI_model}
   \end{center}
\end{figure}

This model in based upon the work of Kucharski {\it et~al.} (see~\cite{Kucharskietal2016}).
This model incorporates a new compartment, exposed, for the human and the mosquito
subpopulation which represents the number of individuals (and mosquitoes)
that are incubating the virus, $E_H$ (and $E_v$ in the case of the vector
population), i.e. where individuals (mosquitoes) are infected but are not able yet
to transmit the virus. The inclusion of such a compartment into the model is due to the
fact that it is known that vector and human populations incubate the virus
for a number of days.  Fig.~\ref{SEIRSEI_model} depicts the structure of this
new model.

All parameters have the same connotation as in the SIR/SI model,
and here $1/\kappa_H$ and $1/\kappa_v$ are the mean latent periods for humans
and mosquitoes population respectively.
The governing equations of the model follow:

\begin{eqnarray}
\frac{dS_H}{dt}(t) & = & \mu_H N_H -\beta_H S_H(t) I_v(t) /N_v - \mu_H S_H(t),
    \label{dsHdt3} \\
\frac{dE_H}{dt}(t)  & = & \beta_H S_H(t) I_v(t) /N_v - \kappa_H E_H(t) - \mu_H E_H(t),
    \label{deHdt3} \\
\frac{dI_H}{dt}(t)  & = & \kappa_H E_H(t) - \gamma_H I_H(t) - \mu_H I_H(t),
    \label{diHdt3} \\
R_H & = & N_H - S_H - I_H - E_H, \label{RH3} \\
\frac{dE_v}{dt}(t)  & = & \beta_v (S_v+E_v) I_H(t) /N_H - \kappa_v E_v(t) - \mu_v E_v(t),
     \label{devdt3} \\
\frac{dI_v}{dt}(t)  & = &  \kappa_v E_v(t) - \mu_v I_v(t), \label{divdt3} \\
S_v & = & N_v - I_v - E_v, \label{Sv3} \\
\frac{dC_i}{dt} & = & p \beta_H S_H(t) I_v(t) / N_v, \label{dciHdt3}
\end{eqnarray}
with the initial conditions in $t = t_0$
\begin{eqnarray}
I_H(t_0) & = & I_{H0} > 0, \label{inicialIH3} \nonumber \\
R_H(t_0) & = & R_{H0} > 0, \label{inicialRH3} \nonumber \\
I_v(t_0) & = & I_{v0} > 0, \label{initialIv3} \nonumber \\
E_H(t_0) & = & E_{H0} > 0, \label{initialEH3} \nonumber \\
E_v(t_0) & = & E_{v0} > 0, \label{initialEv3} \nonumber
\end{eqnarray}
where $I_{H0}$, $R_{H0}$, $I_{v0}$, $E_{H0}$ and $E_{v0}$ are respectively,
the initial number of infected people, the initial number of recovered people,
the initial number of infectious mosquitoes, the initial number of people
incubating the virus, and the initial number of mosquitoes incubating the virus,
respectively.
In equation~\ref{dciHdt3}, $C_i$ accounts for the cumulative infectives and
is a smooth monotone function that is used for identification purposes.

\section{The basic reproductive ratio $R_0$} \label{Sec_R0}
The basic reproductive ratio (see~\cite{Diekmannetal1990, Diekmannetal2010} and
\cite{Heffernanetal2005}), $R_0$, is defined as the expected number of
secondary infections arising from a single individual during his or her
entire infectious period, in a population of susceptibles.
$R_0$ often serves as a threshold parameter that predicts whether an infection
will spread.
Determinig $R_0$ is vital to understand and characterize the dynamics
of the disease. However this crucial parameter is model dependent.

$R_0$ is the dominant eigenvalue  of the so call ``next  generation  matrix''.
It is shown that, if $R_0 < 1$, then the disease free equilibrium is locally
asymptotically stable; whereas if $R_0 > 1$, then it is unstable
(see~\cite{Driesscheetal2002, Driesscheetal2008}).

The basic reproduction number for the SIR model is known to be calculated as
$$R_0 = \frac{\beta}{\mu + \gamma}.$$

The basic reproduction number for the SIR/SI model is known to be calculated
(see ~\cite{Pandey2013}) as:
$$R_0 = \frac{\beta_H \beta_v}{\mu_v (\mu_H + \gamma_H)}.$$

The basic reproduction number for the SEIR/SEI model is known to be calculated
(the dominant  eigenvalue  of the next generation matrix,
see~\cite{Diekmannetal1990, Diekmannetal2010}) as:
$$R_0 = \frac{\beta_H \beta_v \kappa_v}{\mu_v \gamma_v (\kappa_v + \mu_v)}.$$

We study and calculate $R_0$ using different modelling perspectives that can
allow us to draw better conclusions on its validity and range.
Also, this parameter is investigated at different granularity levels:
country wide and state wide.

\section{Results} \label{Sec_Results}
The used data used to validate the models, available on the internet, was
published by Faria {\it et~al.} \cite{FariaEtAl2016, FariaEtAl2016-data} while
gathering information for their research.
The data is partitioned by municipalities in Brazil, thus one may have
different levels of granularity in the visualization of the data.
By aggregation, of cases in different
federal entities one can determine those entities that have the most reported cases.
Table~\ref{entidades_casos} shows the total number of
cases throughout 2015 as reported in the epidemiological data provided.

In table~\ref{entidades_casos} only those entities that report
more than 100 cases in a year are displayed.
One can see that the federal entity that reported  the most
number of cases is Bah\'ia, followed by far by Alagoas, Cear\'a, and Rio
Grande du Norte.  All of these entities are located in the northeastern
part of the country. The northeartern region is characterized by high
temperatures (annual  averages between 20 and 28 $^o$C (68.0 and 82.4 $^o$F),
maxima of around 40 $^o$C (104 $^o$F)). During the months of June and July
temperatures vary  between 12 and 16 $^o$C (53.6 and 60.8 $^o$F) in the coastal
regions, where most cases are reported.

\begin{table}[htp!]
\caption{Reported cases in some entities of Brazil}
\label{entidades_casos}
\begin{center}
\begin{tabular}{|c|c|}
\hline
Entities                  & Reported cases\\
\hline
Bah\'{\i}a (BA)           & 27290\\
Alagoas (AL)              & 497 \\
Cear\'a (CE)                & 416 \\
Rio Grande do Norte (RN)  & 264 \\
Par\'a (PA)               & 155 \\
Esp\'{\i}ritu Santo (ES) & 125 \\
\hline
Total Brazil              & 29639 \\
\hline
\end{tabular}
\end{center}
\end{table}

\begin{figure}[htp!]
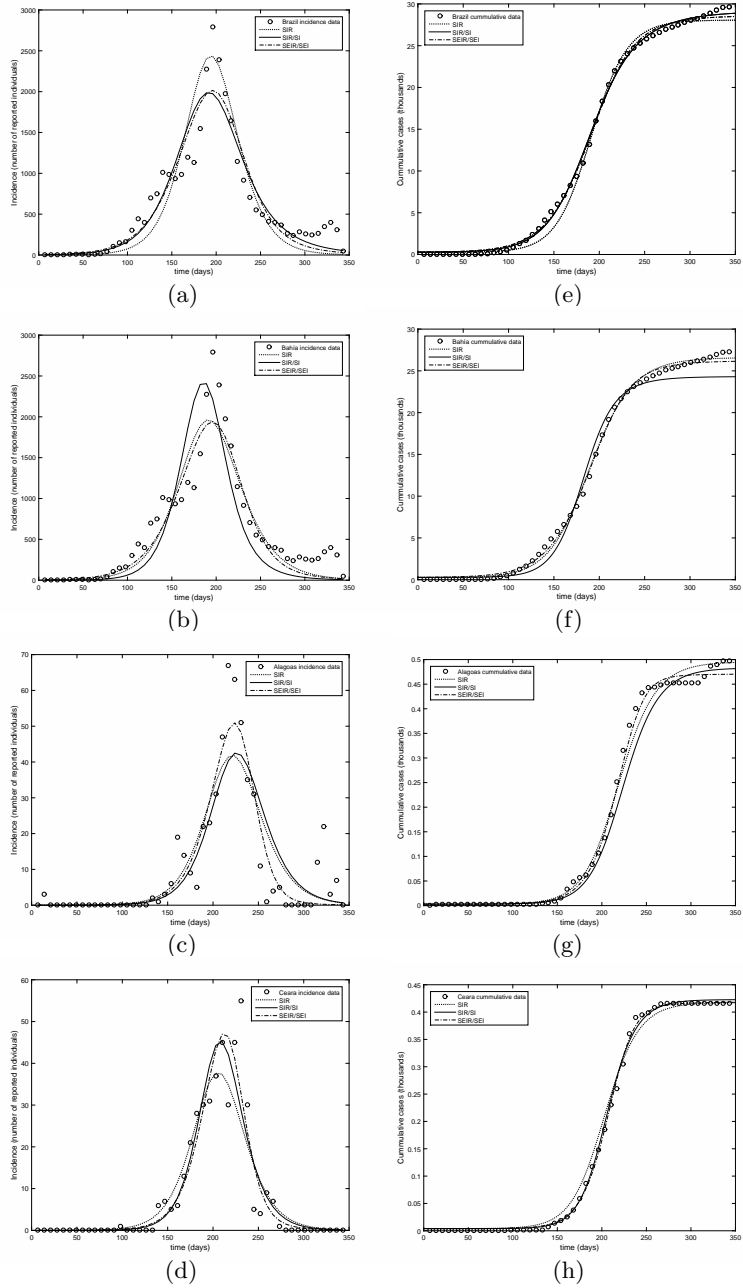

   \begin{center}
   {\includegraphics[width=4.6cm]{IncidenceBrazil} \hspace*{3mm}
   \includegraphics[width=4.6cm]{CummulativeBrazil} \\
   \vspace*{-3mm}
     {\small (a) \hspace*{4.5cm} (e)}   \\
   \vspace*{2mm}
   \includegraphics[width=4.6cm]{IncidenceBahia} \hspace*{3mm}
   \includegraphics[width=4.6cm]{CummulativeBahia} \\
   \vspace*{-3mm}
     {\small (b) \hspace*{4.5cm} (f)}   \\
   \vspace*{2mm}
   \includegraphics[width=4.6cm]{IncidenceAlagoas} \hspace*{3mm}
   \includegraphics[width=4.6cm]{CummulativeAlagoas} \\
   \vspace*{-3mm}
     {\small (c) \hspace*{4.5cm} (g)}   \\
   \vspace*{2mm}
   \includegraphics[width=4.6cm]{IncidenceCeara} \hspace*{3mm}
   \includegraphics[width=4.6cm]{CummulativeCeara}} \\
   \vspace*{-3mm}
     {\small (d) \hspace*{4.5cm} (h)}   \\
   \vspace*{-2mm}
   \caption {Incidence (images (a) to (d)) and cumulative (images (e) to (h))
     data and model results for Brazil ((a) and (e)), Bah\'{\i}a ((b) and (f)),
     Alagoas ((c) and (g)) and Cear\'a ((d) and (h)).
     Points marked with circles in (a) to (d) correspond to incidence data and
     in (e) to (h) correspond to cumulative data.
     Dashed, solid, and dashdotted lines represent SIR, SIR/SI and
     SEIR/SEI models respectively.}
   \label{model-graf-resultados}
   \end{center}
\end{figure}

The incidence data (left panels, images (a) to (d)) and the cummulative
data used for identification purposes (right panels, images (e) to (h))
for the agreggated country data and most reported federal entities
are shown in Fig.~\ref{model-graf-resultados}.


Relevant demographic information taken from the web page of the
Instituto Brasileiro de Geografia e Estat\'{\i}stica (IBGE) (2013)
are given in table~\ref{demografic_data}, where proportion refers
to the proportion of reported cases to total population.

\begin{table}[htp!]
\caption{Relevant demographic data}
\label{demografic_data}
\begin{center}
\begin{tabular}{|c|c|c|c|c|}
\hline
Parameter        & Brazil & Bah\'{\i}a & Alagoas & Cear\'a\\
\hline
Population (MM)  & 207    & 15.15 & 8.86   & 3.32\\
Life expectancy (years)  & 75     & 71.9  & 69.2   & 72.4 \\
Reported cases   & 29639  & 27290 & 416    & 497\\
Proportion       & 1.43e-4 & 1.80e-3 & 4.69e-5 & 1.491e-4\\
\hline
\end{tabular}
\end{center}
\end{table}

The life expectancy was used in all parameter estimations, thus reducing
the number of free parameters to be identified.

A brief description on how to solve the system of ODEs associated to any
of the three models proposed (SIR, SIR/SI and SEIR/SEI) is given in the appendix.
The numerical models, which allow us to calculate the cumulative number of
cases for each model, were implemented in language M of MATLAB.
These functions will be used in the identification process of the unknown
parameters, being 3, 6 and 10 the number of parameters for the SIR, SIR/SI
and SEIR/SEI models respectively (see table~\ref{range-parameters}).

It is important to note that the identification was performed over
nondimensionalized systems.                                           

The problem to be solved for the estimation of the unknown parameters is
the following:
Given a function $f(t)$ which represents the cumulative weekly number of cases
for a period of 49 weeks during 2015, find the parameters $x_i$, $i=1,\dots,n$
such that the answer $\tilde f(t)$ given by the model and the cumulative
incidence $f(t)$ are closed, that is
$$g(x_{min}) = \globalmin_{x \in \Omega} g(x),$$
where $g:\Omega \subset \mathbb{R}^n \rightarrow \mathbb{R}$ defined by
$$g(t) = \int_0 ^T (f(t) - \tilde f(t))^2 dt$$
the mean squared error (MSE),
$\Omega = \prod_{i=1} ^n [a_i,b_i]$ and $T$ is the time corresponding to
the data.

It is important to point out that it is possible to find more than one set
of parameters which satisfy the minimality condition required and related
to the size of $\Omega$. Also relevant is that every time the objective
function needs to be evaluated, one run of the model has to be carried out.

The identification was carried through an exhaustive search procedure
within the range of the different parameters involved in the model,
not being the best available methodology because of the high number of
evaluations of the objective function.

We considered feasible ranges for the parameters of the model
whenever possible.
Thus, according to epidemiological data, previously reported
results \cite{Kucharskietal2016, Castillo-Chavezetal2016}
and some sensibility carried out,
the range considered for the parameters are in table~\ref{range-parameters}.

\begin{table}[htp!]
\caption{Ranges for the estimation of the models unknown parameters.}
\label{range-parameters}
\begin{center}
\begin{tabular}{|c|c|c|c|c|}
\hline
Model & \Lower{SIR} & \Lower{SIR/SI} & \Lower{SEIR/SEI} & reference for \\
Parameter &  &  &  & the reciprocal \\
\hline
$\beta_H$ (day$^{-1}$)& [0.01,0.6] & [0.01,0.6] & [0.1,1.2] & - \\
$\gamma_H$ (day$^{-1}$)& [0.1,0.4] & [0.1,0.5] & [0.1,0.5] & 2-9 days \\
$\beta_v$ (day$^{-1}$)& - & [0.01,0.4] & [0.01,0.3] & - \\
$\mu_v$ (day$^{-1}$)& - & [0.03,0.3] & [0.03,0.3] & 4-30 days \\
$\kappa_H$ (day$^{-1}$)& - & - & [0.083,0.34] & 3-12 days \\
$\kappa_v$ (day$^{-1}$)& - & - & [0.20,0.50] & 2-5 days \\
$I_{v0}$ & - & [1.e-9,5.e-5] & [8.e-7,5.e-5] & - \\
$E_{0}$ & - & - & [8.e-7,9.e-6] & - \\
$E_{v0}$ & - & - & [8.e-7,9.e-6] & - \\
$p$ & [1.e-5,6.e-3] & [1.e-5,6.e-3] & [4.e-5,6.e-3] & - \\
\hline
\end{tabular}
\end{center}
\end{table}

We estimated the total of three unknown parameters for the SIR model
(see table~\ref{EDO_SIR_ajuste}), six parameters for the SIR/SI
model (see table~\ref{EDO_SIRSI_ajuste}), and ten parameters for
the SEIR/SI model (see table~\ref{EDO_SEIRSEI_ajuste}),
using the cumulative incidence suspicious data.
Cumulative incidence is generally smoother than the original incidence data
and thus easier to fit.

The numerical solutions of the models were performed using Matlab and
compared to the cummulative data to obtain a mean squared error (MSE).
Those parameters that produced the smallest errors are reported as the best
identification for the data.
In all cases the basic reproduction number is reported as well as the MSE.

\begin{table}[htp!]
\caption{Estimated parameters for the SIR model.}
\label{EDO_SIR_ajuste}
\begin{center}
\begin{tabular}{|c|c|c|c|c|}
\hline
Parameter & Brazil     & Bah\'{\i}a & Alagoas   & Cear\'a\\
\hline
$\beta_H$  & 0.4340    & 0.1954    & 0.26767    & 0.46124\\
$\gamma_H$ & 0.3816    & 0.1493    & 0.21576    & 0.40667\\
$p$        & 5.7931e-4 & 4.0204e-3 & 4.0912e-4  & 2.0506e-4\\
\hline
$R_0$      & 1.137     & 1.274     & 1.2382     & 1.1336 \\
Error      & 2.2406e-3 & 9.4618e-4 & 4.1092e-3  & 2.4544e-3\\
\hline
\end{tabular}
\end{center}
\end{table}

\begin{table}[htp!]
\caption{Estimated parameters for the SIR/SI model.}
\label{EDO_SIRSI_ajuste}
\begin{center}
\begin{tabular}{|c|c|c|c|c|}
\hline
Parameter  & Brazil    & Bah\'{\i}a & Alagoas  & Cear\'a\\
\hline
$\beta_H$  & 0.49368   & 0.48947   & 0.31889   & 0.54211 \\
$\gamma_H$ & 0.21263   & 0.11      & 0.36333   & 0.31053\\
$\beta_v$  & 0.038947  & 0.037474  & 0.36000   & 0.15474\\
$\mu_v$    & 0.0333    & 0.05      & 0.22      & 0.15667\\
$I_{v0}$   & 3.4526e-5 & 1e-7      & 3e-8      & 1e-9\\
$p$        & 1.5263e-4 & 1.6632e-3 & 2.7778e-4 & 6.8737e-5\\
\hline
$R_0$      & 2.715     & 3.3338    & 1.436     &  1.724\\
Error      & 7.0798e-4 & 6.0564e-3 & 2.9883e-3 & 7.6298e-4\\
\hline
\end{tabular}
\end{center}
\end{table}

Fig.~\ref{model-graf-resultados} overlays the cummulative reported data
points with the simulated cummulative cases with the best set of
parameters for each entity (right panels) and the 3 models investigated.
Although incidence data was not used in the identification purposes,
left panel on the left show an overlay of the incidence data with the
simulated incidence.

\begin{table}[htp!]
\caption{Estimated parameters for the SEIR/SEI model.}
\label{EDO_SEIRSEI_ajuste}
\begin{center}
\begin{tabular}{|c|c|c|c|c|}
\hline
 Parameter & Brazil     & Bah\'{\i}a & Alagoas   & Cear\'a \tabularnewline
\hline
$\beta_H$  & 0.86372    & 0.84      & 0.8       & 1.1 \tabularnewline
$\gamma_H$ & 0.28333    & 0.19667   & 0.11      & 0.305 \tabularnewline
$\beta_v$  & 0.045      & 0.04      & 0.040556  & 0.06 \tabularnewline
$\mu_v$    & 0.03333    & 0.03333   & 0.03333   & 0.03333 \tabularnewline
$\kappa_H$ & 0.083      & 0.083     & 0.083     & 0.083 \tabularnewline
$\kappa_v$ & 0.425      & 0.2       & 0.2       & 0.5 \tabularnewline
$I_{v0}$   & 5.e-5      & 5.e-5     & 8.e-7     & 1.e-8 \tabularnewline
$E_{0}$    & 9.e-6      & 9.e-6     & 8.e-7     & 6.e-7 \tabularnewline
$E_{v0}$   & 4.9e-6     & 6.95e-6   & 8e-7      & 6.e-7 \tabularnewline
$p$        & 0.00014    & 0.00173   & 0.00014   & 0.000046667 \tabularnewline
\hline
$R_0$      & 3.8165     & 4.3936     & 7.5854    & 6.0867 \tabularnewline
Error  & 0.00081171 & 0.00089141 & 0.0015467 & 0.00052088 \tabularnewline
\hline
\end{tabular}
\end{center}
\end{table}

\section{Discussion} \label{Sec_Discussion}
In this work we identified three models for the Zika virus in Brasil.
The models were identified using the aggregated data for the country
and then for 3 federal entities, namely Bahia, Alagoas and Cear\'a, which
were the areas with the greatest number of cases reported.
One aspect worth mentioning is that the mayority of the cases came
from Bah\'{\i}a.  One can see from the results in
tables \ref{EDO_SIRSI_ajuste}, and \ref{EDO_SEIRSEI_ajuste} that the
model parameters are similar for the aggregated data (Brazil) and the
federal region with most numbered cases.
This is so for those models that include the vector within the models,
namely the SIR/SI and SEIR/SEI.  However, this is not maintained in
the more basic SIR model (see Table \ref{EDO_SIR_ajuste}).

For the SIR model, the results for Cear\'a and the aggregated Brazil
cases are similar according to the estimated parameters, while Bah\'{\i}a
and Alagoas seemed tuned into one another.
This result is in agreement with the proportions calculated
in Table \ref{demografic_data}
for Brazil and Cear\'a, but they do not agree when comparing Bah\'{\i}a and
Alagoas since there is a difference in two orders of magnitude.
But the models adjust best to similar contact and infectious rates.
It is noteworthy however that the $R_0$ is within
similar ranges (1.1 - 1.3), see \cite{Pandey2013} with $R_0 = 1.10$ for SIR model,
but far from the ranges reported in other Zika outbreaks,
see \cite{Nishiuraetal2016} ($R_0 = 4.3-5.8$ for the Yap Island
epidemic and $R_0 = 1.8-2.2$ for the French Polynesia epidemic), and
\cite{Wallingaetal2007} ($R_0 = 1.57$ for the SIR model, $R_0 = 1.65$
using a SEIR model, and $R_0 = 1.66$ for a more complicated model, all applied
to influenza A).
One reason could be that the $R_0$ is model dependent and in this case
this very simple model does not include important infectious factor such
as the vector population.  Therefore, we regard this model to be
non-informative and too simplistic for the situation being modelled,
and it was considered for comparison purposes and as a parsimonious model.
When the vector population is included the ranges of the values for
$R_0$ expand and the values are bigger.
For the SIR/SI model the
range is (1.4 - 3.3) and for the SEIR/SEI model the range is (3.8 - 7.5),
this could be due to model dependency of $R_0$.
Although the ranges for the $R_0$ are disjunct, these are consistent with
other reported values in the literature, see
\cite{Castillo-Chavezetal2016} with $R_0 = 4.4$ with 95\% CI $[3.0,6.2]$
and a one standard deviation uncertainty of $0.9$ for the
outbreak of ZIKV that began in 2015 in Barranquilla, Colombia using the SEIR/SEI model,
\cite{Pandey2013} with $R_0 = 1.57$ using a vector-host model, and
\cite{Kucharskietal2016} with $R_0$ ranged for from 2.6 (95\% CI: $[1.7,5.3]$) in
Marquises to $4.8$ (95\% CI: $[3.2,8.4]$) in Moorea for French Polynesia ZIKV outbreak using
the SEIR/SEI model.

Even though the models that included the vector population have more free
parameters to adjust and identify, allowing for greater degrees of freedom,
the errors in the adjustments are similar for all three models used.
Furthermore, all three models follow the data very well as one can see on
the (e) to (h) panels in Fig.~\ref{model-graf-resultados}.
Even though the incidence data was not used during
the identification process, panels (a) to (d) from Figure Fig.~\ref{model-graf-resultados}
overlays the data with the induced incidence curve with the model parameters found.
In the case of Brazil, the SIR model was able to best capture the rise in the
incidence around the peak of infectiousness (day 200), but for the
subregions the models that incorporated the vector population best
captured such rise.

One point to mention is that the Brazil data aggregates information from
varying regions with varying forms and times of outbreak, which makes
the interpretation of the results difficult.

\section*{Acknowledgements}
This research did not receive any specific grant from funding agencies in the public,
commercial, or not-for-profit sectors.


\bibliography{Zika-BuitragoEscalanteVillasana-v1}

\section*{Appendix - Discretizing the models}
In order to exemplify the discretization methodology, the SIR/SI
model (equations \ref{dsHdt2} to \ref{dciHdt2}) will be used.
Given the fact that the human population remains constant, one can
express $R_H$ in terms of the variables $S_H$ and $I_H$, i.e.
$R_H = N_H - S_H - I_H$,
therefore, equation~\ref{RH2} for this model can be discarded
and we can reduce the dimensionality of the system. A similar argument
is true for the case of $S_v$ in equation~\ref{Sv2}, i.e. $S_v = N_v - I_v$.
Finally, the following system of ordinary differential equations with initial
conditions has to be solved.

\begin{eqnarray}
\frac{dS_H}{dt}(t) & = & \mu_H N_H -\beta_H S_H(t) I_v(t) /N_v
     - \mu_H S_H(t), \label{App-dsHdt} \\
\frac{dI_H}{dt}(t)  & = & \beta_H S_H(t) I_v(t) /N_v - \gamma_H I_H(t)
     - \mu_H I_H(t), \label{App-diHdt} \\
R_H & = &  N_H - S_H - I_H \label{App-RH} \\
\frac{dI_v}{dt}(t)  & = & \beta_v S_v(t) I_H(t) /N_H - \mu_v I_v(t),
      \label{App-divdt} \\
S_v & = & N_v - I_v, \label{App-Sv} \\
\frac{dC}{dt} & = & p \beta_H S_H(t) I_v(t) / N_v, \label{App-dciHdt}
\end{eqnarray}
with the initial conditions in $t = t_0$
\begin{eqnarray}
I_H(t_0) & = & I_{H0} > 0, \label{App-inicialIH} \nonumber \\
R_H(t_0) & = & R_{H0} > 0, \label{App-inicialRH} \nonumber \\
I_v(t_0) & = & I_{v0} > 0. \label{App-initialIv} \nonumber
\end{eqnarray}

The following step is the normalization of this system, i.e. with
\begin{align*}
s_H &= S_H/N_H, & i_h &= I_H/N_H, & r_H &= R_H/N_H, \\
i_v &= I_v/N_v, & s_v &= S_v/N_v, & c &= C/N_H,
\end{align*}
the following equivalent system of ODEs arises
\begin{eqnarray}
\frac{ds_H}{dt}(t) & = & \mu_H -\beta_H s_H(t) i_v(t)
     - \mu_H s_H(t), \label{App-dsHdt1} \\
\frac{di_H}{dt}(t)  & = & \beta_H s_H(t) i_v(t) - \gamma_H i_H(t)
     - \mu_H i_H(t), \label{App-diHdt1} \\
r_H & = &  1 - s_H - i_H \label{App-RH1} \\
\frac{di_v}{dt}(t)  & = & \beta_v s_v(t) i_H(t) - \mu_v i_v(t),
      \label{App-divdt1} \\
s_v & = & 1 - i_v, \label{App-Sv1} \\
\frac{dc}{dt} & = & p \beta_H s_H(t) i_v(t). \label{App-dciHdt1}
\end{eqnarray}

The basic idea of any approximation method is to replace the original problem
by another problem that is easier to solve and whose solution is, is in some sense,
close to the solution of the original problem.

Given $M \in \mathbb{N}$, let $\{t_i\}_{1 \leq i \leq M}$ be an uniform
subdivision of the time domain, with mesh length $h = \Delta t = t_{i+1}-t_i$
along the direction $t$.

The following is the forward finite difference for the first order operator
$$\left(\frac{df}{dt}\right)_{i+1} = \frac{f(t_i+h) - f(t_i)}{h} \ \
       {\rm for} \ 1 \leq i \leq M-1$$
where $f$ represents the functions $s_H$, $i_H$, $r_H$, $i_v$, $s_v$ and $c$.

Finally the following set of linear equations arises upon susbtitution of the
forward finite difference operator
\begin{eqnarray}
s_H(t_i+h) & = & s_H(t_i) +\mu_H \, h -\beta_H s_H(t_i) i_v(t_i) \, h
     -\mu_H s_H(t_i) \, h, \label{App-dsHdt2} \\
i_H(t_i+h) & = & i_H(t_i) + \beta_H s_H(t_i) i_v(t_i) \, h
     -\gamma_H i_H(t_i) \, h -\mu_H i_H(t_i) \, h, \label{App-diHdt2} \\
r_H(t_i+h) & = &  1 - s_H(t_i+h) - i_H(t_i+h) \label{App-RH2} \\
i_v(t_i+h) & = & i_v(t_i) +\beta_v s_v(t_i) i_H(t_i) \, h
     -\mu_v i_v(t_i) \, h, \label{App-divdt2} \\
s_v(t_i+h) & = & 1 - i_v(t_i+h), \label{App-Sv2} \\
c(t_i+h) & = & c(t_i) +p \beta_H s_H(t_i) i_v(t_i) \, h, \label{App-dciHdt2}
\end{eqnarray}
for $\ 1 \leq i \leq M-1$.

The numerical model (equations \ref{App-dsHdt2} to \ref{App-dciHdt2})
was implemented using the M languages of MATLAB.

\end{document}